\documentclass[preprint]{aastex}

\usepackage{emulateapj5}

\newcommand{\Spitzer}{{\sl Spitzer}}
\newcommand{\ISO}{{\sl ISO}}

\newcommand{\Msun}{\mbox{$M_{\sun}$}}
\newcommand{\Mearth}{\mbox{$M_{\oplus}$}}
\newcommand{\Lsun}{\mbox{$L_{\sun}$}}

\newcommand{\perone}{\mbox{$^{-1}$}}
\newcommand{\pertwo}{\mbox{$^{-2}$}}

\newcommand{\etal}{et al.}
\newcommand{\eg}{e.g.}
\newcommand{\ie}{i.e.}
\newcommand{\cf}{cf.}
\newcommand{\IRAS}{{\sl IRAS}}

\newcommand{\kms}{\hbox{km~s$^{-1}$}}

\newcommand{\htwo}{{\hbox{H$_2$}}}     

\newcommand{\degs}{\mbox{$^{\circ}$}}

\newcommand{\Mstar}{\mbox{$M_{\star}$}}

\newcommand{\bPic}{\hbox{$\beta$}~Pic}

\slugcomment{{\sl Astrophysical Journal}, accepted February 2004}

\shorttitle{Debris Disks around Nearby Young Stars}
\shortauthors{Liu et al.}

\begin{document}

\title{A Sub-Millimeter Search of Nearby Young Stars for Cold Dust:\\
Discovery of Debris Disks around Two Low-Mass Stars}


\author{\sc Michael C. Liu,\altaffilmark{1} 
            Brenda C. Matthews,\altaffilmark{2,3} 
            Jonathan P. Williams,\altaffilmark{1} 
            Paul G. Kalas\altaffilmark{2}}


\altaffiltext{1}{Institute for Astronomy, University of Hawai`i, 2680
Woodlawn Drive, Honolulu, HI 96822} 

\altaffiltext{2}{Astronomy Department, University of California,
Berkeley, CA 94720}

\altaffiltext{3}{Radio Astronomy Laboratory, University of California,
Berkeley, CA 94720}

\begin{abstract}
We present results from a JCMT/SCUBA 850~\micron\ search for cold dust
around eight nearby young stars belonging to the \bPic\
($t\approx12$~Myr) and the Local Association ($t\approx50$~Myr) moving
groups.
Unlike most past sub-mm studies, our sample was chosen solely on the
basis of stellar age.
Our observations achieve about an order of magnitude greater sensitivity
in dust mass compared to previous work in this age range.
We detected two of the three M~dwarfs in our sample at 850~\micron,
GJ~182 and GJ~803 ($\Mstar \approx 0.5~\Msun$), with inferred dust
masses of only $\approx$0.01--0.03~\Mearth.
GJ~182 may also possess a 25~\micron\ excess, indicative of warm dust in
the inner few~AU of its disk.  
For GJ~803 (AU~Mic; HD~197481), sub-mm mapping finds that the
850~\micron\ emission is unresolved.  A non-detection of the CO~3--2
line indicates the system is gas-poor, and the spectral energy
distribution suggests the presence of a large inner disk hole
($\approx$17~AU = 1.7\arcsec\ in radius for blackbody grains).  These
are possible indications that planets at large separations can form
around M~dwarfs within $\sim$10~Myr.  In a companion paper (Kalas, Liu
\& Matthews 2004), we confirm the existence of a dust disk around GJ~803
using optical coronagraphic imaging.  Given its youthfulness, proximity,
and detectability, the GJ~803 disk will be a valuable system for
studying disk, and perhaps planet, formation in great detail.
Overall, sub-mm measurements of debris disks point to a drop in dust
mass by a factor of $\sim$10$^3$ within the first $\sim$10~Myr, with the
subsequent decline in the masses of sub-mm detected disks consistent
with $t^{-0.5}$ to $t^{-1}$.
\end{abstract}

\keywords{circumstellar matter --- planetary systems: formation ---
planetary systems: protoplanetary disks --- stars: late-type}


\section{Introduction}

Circumstellar disks are common around stars at ages of a few million
years.
Rotationally supported disks are a natural outcome of the star
formation process, because of the need to conserve angular momentum
during the collapse of the natal molecular core.  In addition, disks are
central to current theories of planet formation.  These young disks are
composed of gas and dust, with estimated masses as large as several
percent of the stellar mass.  By virtue of their large optical depths,
they can be easily detected and have been intensively studied.

Over the span of $\sim$5--10 Myr, these primordial disks are thought to
dissipate \citep[e.g.][]{1993prpl.conf..837S, 2001ApJ...553L.153H},
based on the observed decline of IR excesses, which originate from warm
dust in the inner disk.  At older ages, planetesimals and planets
comprise the remaining circumstellar population.
Collisions between the planetesimals produce dust grains, which form a
"second-generation" disk.  The {\sl Infrared Astronomical Satellite}
(\IRAS) discovered the first example of a such dusty debris disk around
Vega \citep{1984ApJ...278L..23A}.  Overall, volume-limited far-IR
surveys have found that about 15\% of main-sequence stars have debris
disks \citep{1993prpl.conf.1253B, 1999ApJ...520..215F,
1999ApJ...510L.131G, 1999A&A...343..496P, 2001A&A...365..545H,
2002A&A...387..285L}.  Such disks have a small amount of dust, little
evidence for gas, low characteristic temperatures indicative of grains
at tens of~AU separations, and a dearth of warm material in their inner
regions \citep[e.g.,][]{2001ARA&A..39..549Z}.  These attributes make
debris disks physically distinct from primordial disks and lead to much
weaker observational signatures.

Loss processes for the dust in debris disks act quickly compared to the
age of the host stars, and the debris must be continually replenished.
Thus, studying debris disks over a wide range of ages can provide
insight into their evolution and physical origin.  Most studies have
focused on stars with ages of $\gtrsim$200~Myr
\citep{1999Natur.401..456H, 2001A&A...365..545H, 2001ApJ...555..932S,
2002A&A...387..285L}, a reflection of the fact that most nearby stars
are old.

The last few years have seen the discovery of many stars which are both
young ($t\approx10$--100~Myr) and close to Earth ($d\lesssim100$~pc).
These stars have largely been found in kinematically associated moving
groups, by combining accurate space motions with evidence for
youthfulness (\eg, Li abundances, X-ray emission, and chromospheric
activity).  They include the \bPic\ ($\approx$8--20~Myr;
\citealp{2001ApJ...562L..87Z}) and the Local Association
($\approx$20--150~Myr; \citealp{1995MNRAS.273..559J,
2001MNRAS.328...45M}) moving groups.  These are important systems for
disk studies as they span an age range which is poorly studied.  In
addition, this timescale is associated with assemblage of terrestrial
planets in our own Solar System \citep[e.g.][]{2000prpl.conf..995W}, so
understanding disk properties at this epoch can illuminate the overall
physical context for planet formation.

Given the cold characteristic temperatures of the dust
($\approx$40--100~K; \eg, \citealp{2000MNRAS.314..702D}), sub-mm
continuum observations of debris disks are valuable.  The emission is
optically thin and hence can be used to estimate the disk masses, with
the caveat that large bodies are missed.  Previous sub-mm studies have
focused on stars already known to have strong IR excesses
\citep{1993ApJ...414..793Z, 1996MNRAS.279..915S, 1998Natur.392..788H,
1998ApJ...506L.133G, 2000MNRAS.313...73S, 2001MNRAS.327..133S,
2003ApJ...582.1141H, 2003AJ....125.3334H}, with the most outstanding
examples being spatially resolved at sub-mm wavelengths (\bPic,
$\epsilon$~Eri, Fomalhaut, Vega).

Much less has been done in the way of "blind" searches of stars in a
given age range to understand their sub-mm disk frequency and
properties.  \citet{2003MNRAS.342..876W} have recently surveyed the
secondary stars of binaries with O and B-type primaries, a sample
originally identified by \citet{1986A&A...156..223L}.  By virtue of the
primary stars being on the main sequence, the systems are
$\lesssim$100~Myr old. Previous searches for dust around this sample had
relatively poor sensitivity \citep{1994A&A...282..123G,
1994AJ....108..661J, 1995ApJ...440L..89R}. Of particular interest is the
suggestion by Wyatt~\etal\ that some stars may harbor very cold
($\lesssim$40~K) disks, which would be missed by previous far-IR
searches and only detected in the sub-mm.  However, one limitation of
the Lindroos sample is that most of the systems are fairly distant,
which impacts the resulting disk sensitivity of sub-mm observations.

In this Letter, we present an 850~\micron\ search for cold dust around
young stars in the \bPic\ and Local Association moving groups.  Unlike
most previous sub-mm studies, our sample was chosen solely on the basis
of stellar age.  Our sample includes three M~dwarfs (GJ~182, 799 and
803) which \citet{1999ApJ...520L.123B} identified as the youngest of the
nearby M~dwarfs, based on the stars' pre-main sequence location on the
color-magnitude diagram.  The majority of debris disks from IR studies
have been identified around A, F, and G-type dwarfs, simply due to
observational sensitivity bias (\eg, compilations by
\citealp{2003MNRAS.345.1212G} and \citealp{2003astro.ph.11546Z}).  The
loss processes for circumstellar dust depend on the host star luminosity
and gravity.  Therefore, studying debris disks over a range of stellar
masses may also advance our understanding of this phenomenon.


\section{Observations}

Sub-mm observations were made from December~2002 to November~2003 using
the SCUBA bolometer array \citep{1999MNRAS.303..659H} at the 15-m James
Clerk Maxwell Telescope (JCMT) on Mauna Kea, Hawaii.\footnote{The JCMT
is operated by the Joint Astronomy Centre in Hilo, Hawaii on behalf of
the parent organizations Particle Physics and Astronomy Research Council
in the United Kingdom, the National Research Council of Canada and The
Netherlands Organization for Scientific Research.}  Zenith optical
depths ranged from 0.25 to 0.35 at 850~\micron.  GJ~803 was first
observed on October~5,~2003 and well-detected at 850~\micron;
additional 450 \micron\ data were obtained on November~12--13,~2003 in
dry, stable conditions, with 450~\micron\ zenith optical depths ranging
from 1.37 to~1.47.  We used the photometry mode to maximize observing
efficiency, with integration times ranging from 18~to 96~min on-source.
The data were reduced using Starlink software and standard reduction
techniques \citep[see][]{2002MNRAS.336...14J}.  Pointing was checked
approximately once per hour and was accurate to better than 2\arcsec\
rms over each night.  Flux calibration factors were derived from
observations of Uranus or Neptune when available, but secondary
calibrators (CRL~618 or CRL~2688) were also used.

We also observed GJ~803 with SCUBA on November~12--13, 2003 in
jiggle-mapping mode.  The weather was stable during the observations
with 850~\micron\ zenith optical depths ranging from 0.26 to 0.29.
Pointing was checked every 1--2 hours using Neptune and was accurate to
an rms error of 2\arcsec.  Flux calibrations were performed by observing
Uranus and Neptune at the start of each night and were consistent
between the two nights to within 10\%.  A total of 24~individual maps
were made at a range of image rotation angles on the array.
GJ~803 was well-detected though the emission is not extended at the
signal-to-noise level (S/N$\approx$8) and angular resolution (14\arcsec
= 140~AU) of the map.  The integrated 850~\micron\ flux in the map is
consistent with earlier measurements made in photometry mode.  The
emission is also consistent with being centered on the star given the
pointing uncertainties.

A sensitive search for CO~$J=$~3--2 line emission from GJ~803 was
conducted with the B3~dual-channel heterodyne receiver and digital
autocorrelating spectrometer (DAS) backend on the JCMT on November~5,
2003.  Observations were performed in beam-switching mode with a
60\arcsec\ throw. Weather conditions were good, and the system
temperature ranged from 400~to 450~K over the elevation range of the
observations (30--40\degs).
A total of 200~min of on-source integration was obtained.  No signal was
detected to a 1$\sigma$~level of 13~mK per 1~\kms\ velocity channel.


\section{Results}

Table~1 presents our sample and SCUBA photometry.  We detected GJ~182
and GJ~803 at 850~\micron, with a possible (formally 2$\sigma$)
detection of GJ~803 at 450~\micron.  
GJ~803 is also detected at 60~\micron\ in the \IRAS\ Faint Source
Catalog (FSC) \citep{1991A&A...244..433M, 2002AJ....124..514S}.  In
addition, we searched for all our targets in the \IRAS\ dataset using
the SCANPI algorithm \citep{IRAS-scanpi}, via the Infrared Processing
and Analysis Center (IPAC) web site.  This search revealed 25~\micron\
(color-corrected) fluxes of $65\pm14$ and $157\pm26$~mJy for GJ~182 and
GJ~803, respectively, which were not included in the FSC.

\begin{figure}
\vskip -0.25in
\centerline{\includegraphics[width=3.5in,angle=90]{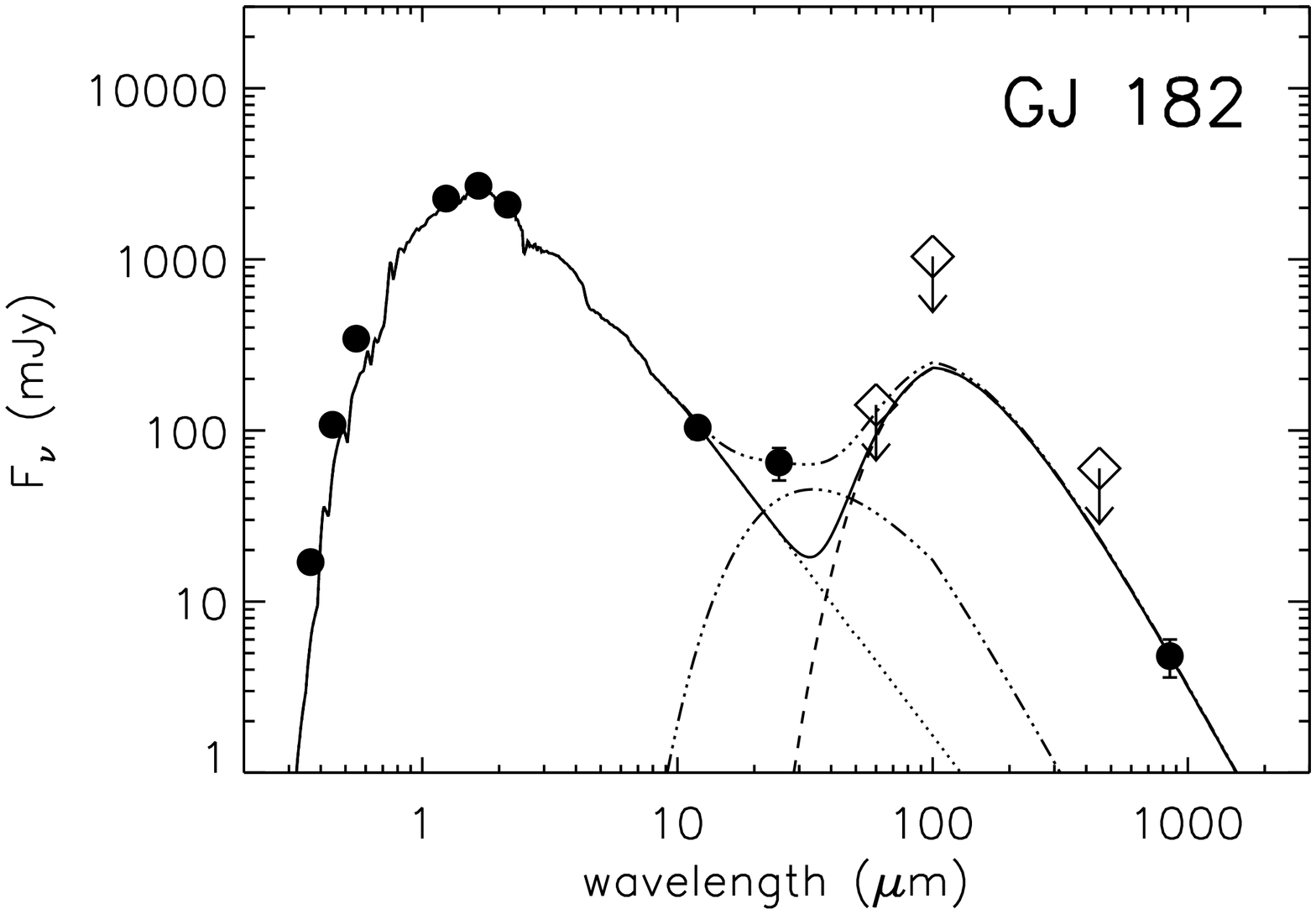}}
\centerline{\includegraphics[width=3.5in,angle=90]{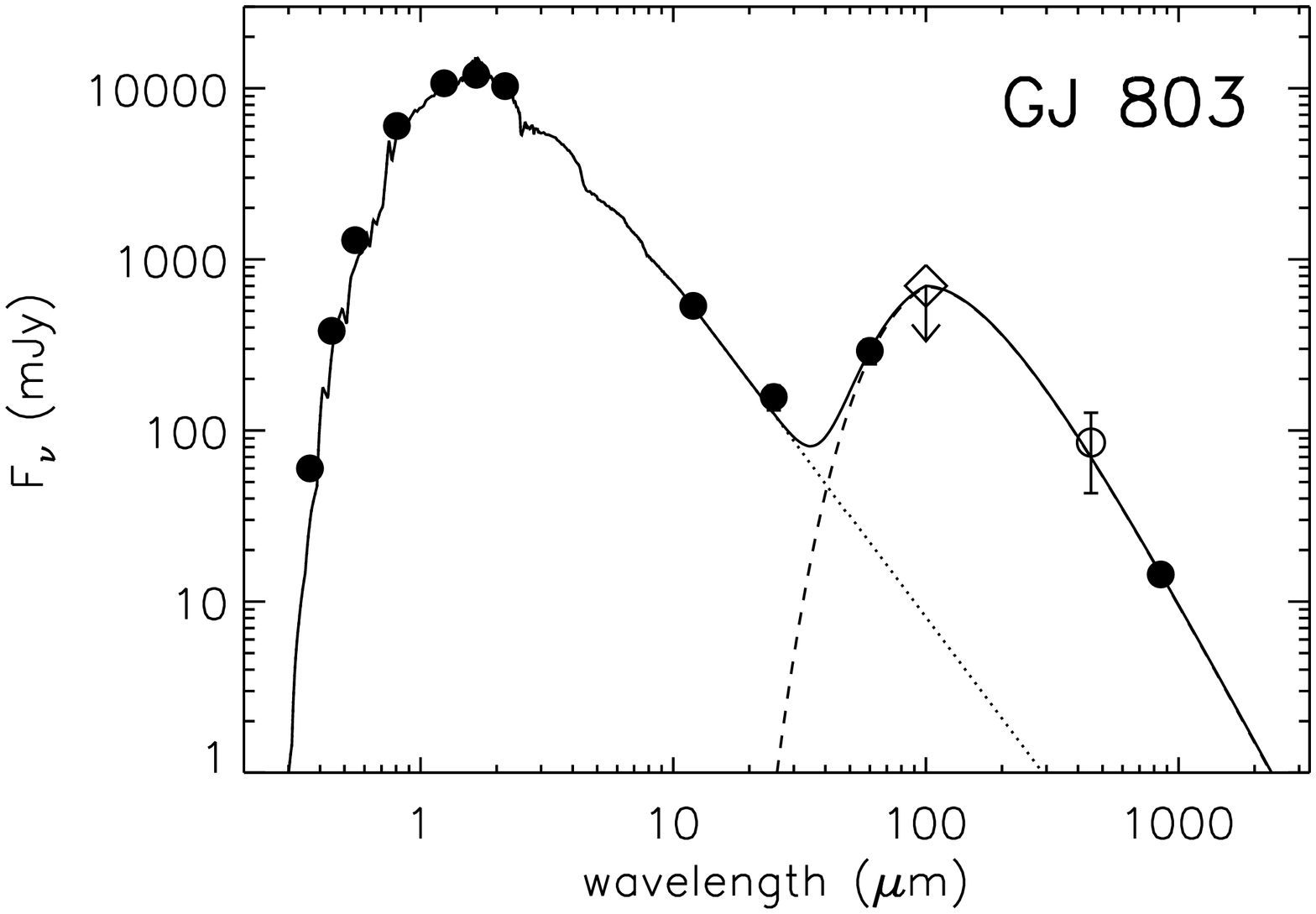}}
\vskip -2ex
\caption{\normalsize Observed SEDs of our stars detected by SCUBA.
Detections are plotted as filled circles and non-detections as open
symbols with arrows.  The possible detection of GJ~803 at 450~\micron\
is shown as an open circle.  Error bars are shown when they are larger
than the symbol size.  The data are compared to the sum of a theoretical
stellar spectrum ({\em dotted line}) and a modified blackbody with
$T=40$~K and $\beta=0.8$ which fits the $\ge$60~\micron\ detections
({\em dashed line}).  For GJ~182, we add a 150~K component to account
for the possible 25~\micron\ excess ({\sl dotted-dashed line}). See text
for details (\S~3 and \S4).  The optical data (0.3--0.8~\micron) are
from SIMBAD and the {\sl Hipparcos} catalog.  The IR data
(1--100~\micron) are from the 2MASS catalog and \IRAS\ color-corrected
Faint Source Catalog and SCANPI photometry.  The sub-mm data (450 and
850~\micron) are from this paper.  The stellar spectra come from the
NextGen models by \citet{2000ApJ...540.1005A, 2001ApJ...556..357A} and
are normalized to the observed $K$-band magnitude. \label{plot-sed}}
\end{figure}

For GJ~803, Figure~\ref{plot-sed} shows the 60~and 850~\micron\ excess
emission modeled with a single temperature fit.  (The 25~\micron\ flux
is consistent within 1$\sigma$ with being purely photospheric in
origin.)  For the fit, we use a modified blackbody, where the emissivity
is constant for $\lambda<100~\micron$ and follows $\lambda^{-\beta}$ for
longer wavelengths.  The spectral energy distributions (SEDs) of debris
disks with more extensive multi-wavelength data are well described by
this prescription, with $\beta=$~1.1~to~0.5 \citep{2000MNRAS.314..702D}.
For this range of $\beta$, the GJ~803 data are fit with $T=35-45$~K,
with $T=40\pm2$~K for a nominal $\beta=0.8$, where the uncertainty is
the formal 1$\sigma$ error in the fit.  (Fitting a pure blackbody to the
data gives $62\pm5$~K.)  Note that the \IRAS\ non-detection at
100~\micron\ is a good match to $\beta=0.8$ and rules out $\beta\gtrsim1$.
With slightly greater sensitivity, \IRAS\ should have detected this
source.  In addition, the marginal detection at 450~\micron\ agrees well
with the $\beta$=0.8 fit, but is inconsistent with a pure blackbody.
The integrated fractional dust luminosity $L_{dust}/L_{star}$ is
$6.1\times10^{-4}$ for the nominal $\beta=0.8$, $T=40$~K model.

For GJ~182, the excess emission at 25~\micron\ and 850~\micron\ cannot
be fit with a single modified blackbody that also satisfies the \IRAS\
non-detections.  Ignoring the 25~\micron\ flux for the moment, a 40~K,
$\beta=0.8$ modified blackbody, like that which fits the GJ~803 SED, is
consistent with the 60, 100, and 850~\micron\ data
(Figure~\ref{plot-sed}).
A higher dust temperature would violate the 60~\micron\ upper limit.  Of
course, a lower dust temperature would also be consistent with the
850~\micron\ detection.  Firmer constraints on the dust temperatures
await more sensitive IR/sub-mm measurements.

The mid-IR SEDs are sensitive to the presence of warm dust in the inner
disk regions.  GJ~182 shows a strong 25~\micron\ excess (but see \S~4.2).
If we add a 150--200~K component which emits as a $\beta=0.8$ modified
blackbody, a dust mass of $(4-12)\times10^{-5}$~\Mearth\ is needed to
account for the 25~\micron\ excess, or 0.3--0.9\% of the T=40~K dust
component which satisfies the 850~\micron\ flux (Figure~\ref{plot-sed}).
For GJ~803, the large mid-IR dip indicates an absence of warm dust in
the inner regions.  If we add a warm component which emits as a
$\beta=0.8$ modified blackbody, only a very small amount of 150--200~K
dust is permitted, since the \IRAS\ 25~\micron\ flux is consistent with
being photospheric.  Taking the marginal 25~\micron\ excess at face
value gives a warm dust mass of $(3-9)\times10^{-6}$~\Mearth, or
0.03--0.08\% of the T=40~K dust component.

For the entire sample, we compute dust masses from the 850~\micron\
fluxes in the standard fashion, assuming optically thin emission
characterized by a single temperature:
\begin{equation}
M_{dust} = {{F_\nu d^2}\over{\kappa_\nu B_\nu(T)}}
\end{equation}
where $F_\nu$ is the flux density, $d$ is distance, $\kappa_\nu$ is the
dust opacity at the observing frequency, and $B_\nu$ is the Planck
function for a dust temperature $T$.  We adopt a dust opacity of
1.7~cm$^2$~g\perone, in agreement with past studies
\citep{1993ApJ...414..793Z, 1998ApJ...506L.133G, 1998Natur.392..788H,
2001MNRAS.327..133S, 2003MNRAS.342..876W, 2003astro.ph.11583W,
2003astro.ph.11593S}.  This value is on the upper end of the
0.4--1.7~cm$^2$~g\perone\ range discussed by
\citet{1994ApJ...421..615P}.  For non-detections, we use 3$\sigma$ upper
limits on the 850~\micron\ flux and assume $T=30-100$~K, in agreement
with previously detected debris disks \citep{2000MNRAS.314..702D}.  For
GJ~803, we adopt the 40~K temperature from the SED fitting.  The
calculated dust masses are presented in Table~1.

It is unlikely that our detections are due to background galaxies.
\citet{2002MNRAS.331..817S} estimate a surface density of about
500~objects per square degree brighter than an 850~\micron\ flux of
6~mJy, about the 3$\sigma$ sensitivity of our survey.  The probability
of any background objects to be within an angular distance $\theta$ of a
target is $1-\exp(-\pi \theta^2 \Sigma)$, where $\Sigma$ is the surface
density of background objects above a specified flux level.  We compute
the ensemble probability of detecting any background objects in SCUBA's
central bolometer for our entire sample of 8~objects, accounting for the
fact that we mapped GJ~803.  This gives a 4\% probability that
background sources would produce at least one detection in our survey.

The GJ~803 system appears to have very little molecular gas.  The
CO~3--2 intensity is $-45 \pm 49$~mK~\kms\ integrated over $\pm$7~\kms\
about the stellar velocity.  (This velocity range corresponds to the
maximum orbital speed for a disk viewed at an inclination of 80\degs\
with the inner disk edge of 17~AU derived from the SED fit.)  Assuming
the gas is optically thin and in thermal equilibrium with 40~K dust, the
3$\sigma$ upper limit implies a CO~column density of
$6.2\times10^{13}$~cm\pertwo.  Temperatures as warm as $\approx$150~K,
\eg, if the gas were located in the inner few AU, would give inferred CO
column densities of a factor of~2 larger.  
Determining an upper limit on the total gas mass is uncertain, since
this is dominated by \htwo\ and the \htwo~to~CO conversion is highly
uncertain.  CO may freeze-out onto grains, making it a poor tracer of
the total disk mass.  Also, photoionization may affect the \htwo~to~CO
conversion factor.
To attempt to account for photoionization by the interstellar UV~field,
we refer to calculations by \citet{1988ApJ...334..771V}, slightly
extrapolated to lower CO column densities, and adopt an \htwo~to~CO
abundance ratio of 10$^{-7}$ (\cf, normal abundance ratio of 10$^{-4}$).
This results in an \htwo\ mass of 1.3~\Mearth, if we ignore the
possibility of CO freeze-out.  Overall, the gas non-detection limits for
GJ~803 are comparable to those for older solar-type stars
\citep{1995MNRAS.277L..25D, 2000MNRAS.312L...1G} and rules out the
possibility of GJ~803 having a gas-rich disk comparable to those
detected around younger stars \citep{1995Natur.373..494Z,
2000Icar..143..155G}.


\section{Discussion}

\subsection{Sub-Millimeter Evolution of Debris Disks}

\begin{figure}
\centerline{\includegraphics[width=4.5in,angle=90]{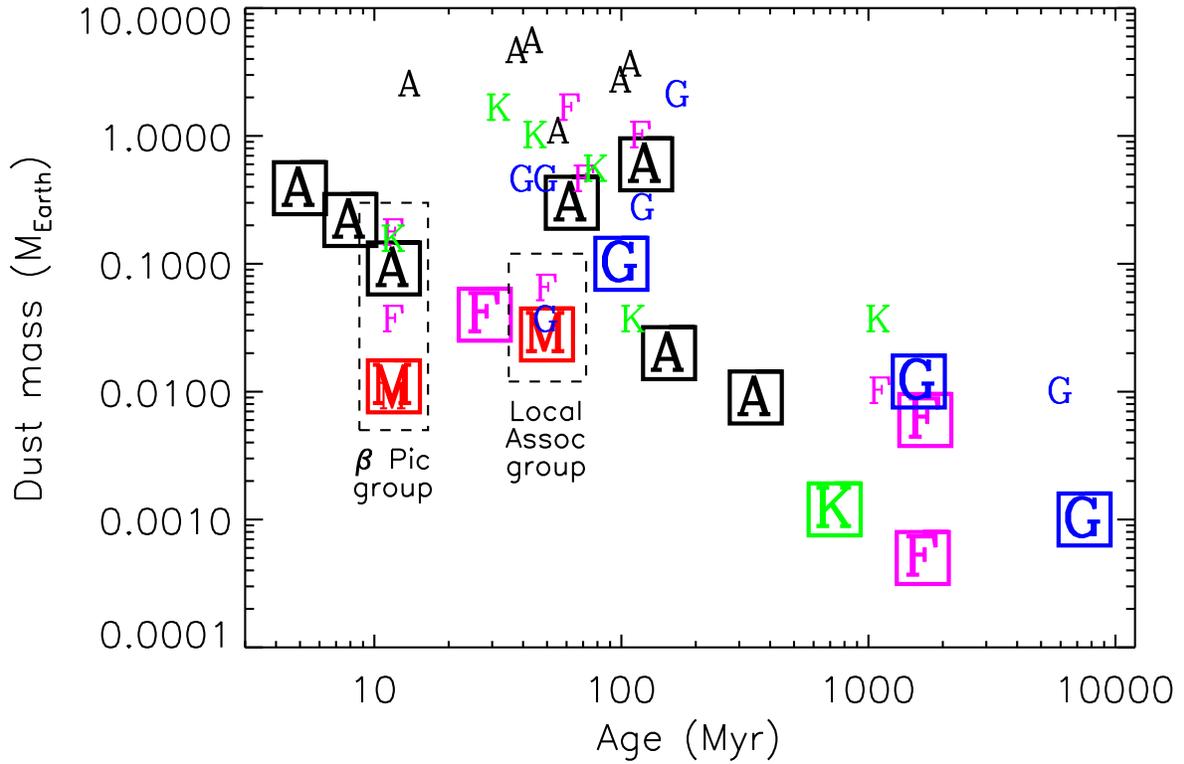}}
\vskip -2ex
\caption{\normalsize Debris disks with dust masses derived from
sub-millimeter data.  Detections are shown in boxes, where the letters
indicate the spectral type.  GJ~182 and GJ~803 are the M~dwarfs detected
in this paper.  Non-detections are shown as smaller letters without
boxes, assuming a dust temperature of 30~K.  The GJ~799
non-detection is obscured behind the M~dwarf detection of GJ~803 at
12~Myr.  In addition to this paper, data come from compilations by
\citet{2003MNRAS.342..876W} and \citet{2003astro.ph.11593S} along with
the young G~star debris disk from \citet{2003astro.ph.11583W}.  The
quoted age uncertainties range from $\approx$25\% to a factor of
$\approx$2.\label{agetrend}}
\end{figure}

Figure~\ref{agetrend} summarizes the known debris disk mass estimates
based on sub-mm observations from our work and the published literature.
Our SCUBA survey adds a significant number of stars of
$\approx$10--50~Myr.  Our raw 850~\micron\ sensitivity (median rms of
1.9~mJy) is poorer than the \citet{2003MNRAS.342..876W} survey of
Lindroos binaries (median rms of 1.6~mJy).  But given the much closer
distances of our targets, our dust mass sensitivity is better by around
an order of magnitude.  In particular, Figure~\ref{agetrend} shows the
following:

\begin{enumerate}

\item Compared to primordial disks around T~Tauri and Herbig~Ae/Be stars
which have dust masses of $\approx$10--300~\Mearth\
\citep[e.g.][]{1995ApJ...439..288O, 2000prpl.conf..559N}, the dust
masses for the detected young stars are $\sim$10$^3$ smaller.  This
points to very rapid evolution of the circumstellar dust mass within the
first $\sim$10~Myr, likely arising from grain removal (\eg, via
accretion onto the central star) and/or grain growth into larger bodies,
which would greatly diminish the sub-mm/mm emission.

\item The upper envelope for disk masses around A~stars exceeds that for
later-type stars.  This may reflect the distribution of primordial disk
masses; for T~Tauri stars a similar trend is seen, where the upper
envelope of primordial disk masses is larger for more massive stars
\citep{2003astro.ph..4184N}.

\item For stars of $\approx10-100$~Myr old, nearly all the detected
disks at a given spectral type (\ie, stellar mass) are less massive than
the upper limits on non-detections.  This highlights the fact that
existing sub-mm measurements are only sensitive to the most massive of
the young debris disks.

\item The apparent correlation of dust mass with stellar age arises from
observational bias: the young stars detected in the sub-mm are at larger
distances than the old stars.  However, the decline in the upper envelope
of disk masses with age is probably a real effect.  While there are no
published sub-mm data in the upper right of Figure~\ref{agetrend} (\ie,
old stars with massive disks), the existing sub-mm detections of nearby,
old stars come from sources known to possess large IR excesses in \IRAS\
and \ISO\ data.  Any old stars with even more massive disks would have
been detected by these IR surveys, unless the dust was unprecedentedly
cold.  Based on the data in Figure~\ref{agetrend}, the masses of sub-mm
detected disks evolve roughly as $M_{dust}\sim t^{-0.5}\ {\rm to}\
t^{-1}$, with a simple unweighted fit giving $M_{dust} \propto
t^{-(0.7\pm0.2)}$.

\end{enumerate}


\subsection{Debris Disks and Planets around Low-Mass Stars}

GJ~182 and GJ~803 are the first M-star debris disks detected at
sub-millimeter wavelengths.\footnote{The young M3~dwarf Hen 3-600 also
shows an IR excess \citep{1989ApJ...343L..61D}.  This binary star is a
member of the $\approx$10~Myr old TW Hydrae Association
\citep{2001ApJ...549L.233Z}.  Hen~3-600 has a very large dust optical
depth, a significant 10--20~\micron\ excess \citep{1999ApJ...520L..41J},
and active gas accretion \citep{2000ApJ...545L.141M}.  These
characteristics are distinct from debris disks and indicate that the
Hen~3-600 disk is primordial or else in a transitional state.}  The dust
lifetime is short compared to the stars' ages
\citep[e.g.][]{1993prpl.conf.1253B}.  Therefore, the dust must be
replenished by collisions between planetesimals, either in situ or else
from regions farther out.  Note that for these low luminosity stars
(0.1--0.2~\Lsun), radiation pressure on the grains is negligible
compared to the star's gravity, unlike debris disks around A~stars, \eg,
\bPic\ where micron-sized and smaller grains are expelled
\citep{1988ApJ...335L..79A}.

Our work shows that most debris disks around low-mass stars probably lie
at or below the sensitivity limits of the \IRAS\ survey and JCMT/SCUBA
observations.  GJ~182 is just at the practical limit for \IRAS\ and
SCUBA detection.  GJ~803 is the nearest object in our sample, and among
the nearest known young stars, which makes detection of its very tenuous
dust mass feasible.  The one undetected M~dwarf in our sample, GJ~799,
is just as close and young as GJ~803.  GJ~799 is known to be a 3\arcsec\
binary of comparable magnitude \citep{2001AJ....122.3466M}, which may
impact its dust content.  In addition, its M4.5 spectral type
corresponds to $\Mstar\approx$0.10--0.15~\Msun, about 3--4$\times$
smaller than for GJ~182 and GJ~803 (\eg, see models presented in
\citealp{1997AJ....113.1733H} and \citealp{1999ApJ...525..466L}).
Hence, the non-detection may simply be due to insufficient sensitivity;
our JCMT data for GJ~799 would have to be $\approx$4$\times$ deeper in
order to reach the same $M_{dust}/M_{star}$ as for the two detected
M~dwarfs.  Primordial disks are common around low mass stars
\citep{2001AJ....121.2065H} and even brown dwarfs
\citep{2002astro.ph.10523L} at ages of a few~Myr, and these systems may
very well generate debris disks as they age.  The {\sl Spitzer Space
Telescope} will offer the requisite sensitivity for detecting and
studying such disks.

For GJ~182, the 25~\micron\ excess suggests that dust resides in the
inner regions of its disk: 150--200~K blackbody grains would lie at
1--2~AU from the star.  Given the sparse sampling of the current SED,
one cannot determine if the dust distribution is continuous throughout
the disk, or if there is a gap in the middle regions of the disk (\eg,
from planet formation) between the 25~\micron\ and 850~\micron\ emitting
dust.  However, as shown by \citet{2002AJ....124..514S}, single band
excesses near the sensitivity limit of the \IRAS\ catalog may not be
reliable.  Though the SCANPI detection of GJ~182 has S/N~=~4.6, higher
than the S/N~=~2--4 objects examined by Song \etal, for now we treat the
25~\micron\ excess of GJ~182 as tentative.  More detailed SED
measurements, in particular with \Spitzer, would be invaluable to
characterize this star's disk.


The absence of warm dust in the inner regions of GJ~803 (and perhaps
GJ~182, if the 25~\micron\ detection is spurious) is naturally explained
by an unseen inner companion.  The 40~K temperature characteristic of
the SED is too cold to be explained by sublimation or melting of icy
grains ($\approx$100-170~K).  Without an inner companion, the effect of
Poynting-Robertson drag would cause grains to spiral inward and produce
significant mid-IR emission.  Such inner holes are typical for debris
disk systems.  Indeed, the few debris disks with resolved dust emission
show morphologies highly suggestive of the dynamical influence of an
unseen companion at mean-motion resonances \citep{1998Natur.392..788H,
1998ApJ...506L.133G, 2000ApJ...537L.147O, 2001ApJ...560L.181K,
2002ApJ...569L.115W, 2002ApJ...578L.149Q, 2003ApJ...584L..27W,
2003ApJ...588.1110K, 2003ApJ...597..566T}.

The 40~K characteristic temperature from the SED fitting means that
blackbody grains around GJ~803 would be at a distance of 17~AU, or an
impressively large angular separation of 1.7\arcsec.  (For GJ~182, 17~AU
corresponds to only 0.6\arcsec.)  This provides only a rough guide as to
the angular separation of any interior companion, since it is difficult
to use the SEDs of debris disks to predict the spatial dust distribution
\citep{2003astro.ph.11593S}.
Shallow IR adaptive optics imaging from the Keck Observatory with
0.05\arcsec\ resolution shows no stellar or massive brown dwarf
companion, suggesting that an even lower mass companion is present.  To
date, GJ~876 is the only M~dwarf known to have extrasolar planets
\citep{1998ApJ...505L.147M, 2001ApJ...556..296M}; its two planets have
semi-major axes of 0.13 and 0.21~AU.

Thus the GJ~803 inner disk hole may indicate that planets can exist at
larger separations around M~dwarfs than known so far and that such
planets can form within $\approx$10~Myr.  
Furthermore, given the very young age of the system, a planetary
companion is predicted to have significant thermal emission
\citep[e.g.][]{1997ApJ...491..856B} and hence might be detectable with
deeper AO imaging.  M~dwarfs are by far the most numerous type of star,
so a better understanding of their planet-bearing frequency is needed
for a complete census of the extrasolar planet population.

Follow-up coronagraphic imaging by \citet{klm03} finds that GJ~803 has a
large, nearly edge-on disk seen in scattered optical light.  This disk
is detected from 5--21\arcsec\ (50--210~AU), though the inner and outer
extent are not well-constrained.  Our sub-mm imaging finds that the
850~\micron\ dust emission is unresolved.  This is entirely compatible
with the large optical disk, given that the grains responsible for
scattering may be very cold and/or small, and hence poor sub-mm
emitters.  Deeper sub-mm mapping is needed for stronger constraints on
any extended emission.
In general, the proximity and youthfulness of GJ~803, combined with the
high sub-mm and optical detectability of its disk, make this system an
excellent opportunity to study disk evolution, and perhaps planet
formation, in great detail.


The onset and duration of the debris disk phenomenon remain open
questions.  As discussed by \citet{2003MNRAS.345.1212G}, the disk masses
might slowly decay over time, or perhaps the mass distribution is
bimodal, reflecting ``on'' and ``off'' states for detectability.  The
GJ~803 debris disk offers some insight into this issue.
Using the {\sl Hipparcos} catalog, which is nearly complete for
early-type M~dwarfs (M0--M2) within 10~pc, \citet{2002AJ....124..514S}
found GJ~803 was the only object out of a sample of 152~M~dwarfs to
possess an IR excess in the \IRAS\ FSC.  
This is not due to detection bias, as there are many early-type M~dwarfs
which have closer distances than GJ~803.
GJ~803 is also known to be among the youngest of the nearby M~dwarfs
\citep{1999ApJ...520L.123B, 2001ApJ...562L..87Z}.
If the appearance of debris disks was an intermittent event, Song
\etal\ should have detected older M~dwarfs with IR excesses at both
60~and 100~\micron, if their disks were like the GJ~803 one.  
Either these older M~dwarfs do not have disks, or their disks are much
less massive than GJ~803.
Therefore, at least for the case of M~dwarfs, the GJ~803 system suggests
that the time evolution of debris disks is driven primarily by age, with
the youngest stars having the largest disk masses.


\acknowledgments

We thank Tom Chester for discussions about \IRAS\ SCANPI processing.
This research has made use of the NASA/IPAC Infrared Science Archive
(IRSA), the SIMBAD database, and the Washington Double Star Catalog
maintained at the U.S. Naval Observatory.  We are grateful for support
from NASA grant HST-HF-01152.01 (MCL), NSF grant AST-0228963 (BCM), NSF
grant AST-0324328 (JPW), and NASA Origins Program grant NAG5-11769
(PGK).

\clearpage


\clearpage

\begin{deluxetable}{llccccccc}
\tablecaption{JCMT/SCUBA Observations \label{sample}}
\tabletypesize{\small}
\tablewidth{0pt}
\tablecolumns{2}
\tablehead{
  \colhead{Object} &
  \colhead{SpT\tablenotemark{(a)}} &
  \colhead{$d$\tablenotemark{(a)}} &
  \colhead{$\lambda$} &
  \colhead{$F_\nu$\tablenotemark{(b)}} &
  \colhead{$M_{dust}$\tablenotemark{(b)}} \\
  \colhead{} &
  \colhead{} &
  \colhead{(pc)} &
  \colhead{(\micron)} &
  \colhead{(mJy)} &
  \colhead{(\Mearth)} 
}

\startdata

%
%
\hline
\sidehead{\bPic\ Group ($12_{-4}^{+8}$ Myr)} \hline
HD 35850\tablenotemark{(c)}& F7V    & 26.8 $\pm$ 0.6\phn & 850 &  $<$4.5     & $<$0.01--0.04 \\
HD 199143          	   & F8V    & 47.7 $\pm$ 2.3\phn & 850 &  $<$7.5     & $<$0.05--0.19\\
HD 358623 (AZ Cap) 	   & K7Ve   & 47.7 $\pm$ 2.3\phn & 850 &  $<$6.9     & $<$0.04--0.17\\
GJ 803 (AU Mic)    	   & M1e    & 9.94 $\pm$ 0.13    & 850 & 14.4 $\pm$ 1.8 & 0.011 \\
                   	   &        &                    & 450 & \phn\phn85 $\pm$ 42\phn      &      \\
GJ 799 (AT Mic)    	   & M4.5e  & 10.2 $\pm$ 0.5\phn & 850 &  $<$9.0     & $<$0.003--0.010 \\

& & & & & \\
\hline
\sidehead{Local Association Group ($50_{-30}^{+100}$~Myr)} \hline
EK Dra (HD 129333) 	   & F8     & 33.9 $\pm$ 0.7\phn & 850 &  $<$5.7     & $<$0.02--0.07 \\
HD 77407  	   	   & G0     & 30.1 $\pm$ 0.8\phn & 850 &  $<$4.5     & $<$0.01--0.04 \\
GJ 182\tablenotemark{(d)}  & M0.5Ve & 26.7 $\pm$ 1.8\phn & 850 & \phn4.8 $\pm$ 1.2     &  0.007--0.028 \\
                   	   &        &                    & 450 &  $<$60\phn  & \\
\enddata

\tablenotetext{a}{From SIMBAD and the {\sl Hipparcos} catalog.}

\tablenotetext{b}{For non-detections, we use the 3$\sigma$ upper limits
on the 850~\micron\ flux and assume $T=30-100$~K.  For GJ~803, the dust
mass is computed assuming $T=40$~K, derived from the SED fitting.}

\tablenotetext{c}{\citet{2000PhDT........17S} report a possible IR
excesses based on \ISO\ data, but \citet{2000A&A...357..533D} suggest
the excess may arise from extended cirrus, not from the star itself.}

\tablenotetext{d}{\citet{2001MNRAS.328...45M} show that GJ~182 shares
kinematics with the higher mass Local Association stars, which have
estimated ages of $\approx$50~Myr based on stellar activity and Li
abundances.  However, placing GJ~182 on an HR diagram suggests a younger
age of $\approx$10--20~Myr \citep{1998A&A...335..218F,
1999ApJ...520L.123B}.}

\end{deluxetable}

\end{document}